\documentclass[12pt]{article}
\usepackage{a4wide}
\usepackage{amssymb}
\usepackage{hyperref}
\begin{document}
{\renewcommand{\thefootnote}{\fnsymbol{footnote}}
\begin{center}
{\LARGE  Cosmic tangle:\\ Loop quantum cosmology and CMB anomalies}

\vspace{1.5em}

Martin Bojowald\\
\vspace{0.5em}
Institute for Gravitation and the Cosmos,\\
The Pennsylvania State
University,\\
104 Davey Lab, University Park, PA 16802, USA\\
\vspace{1.5em}
\end{center}
}

\setcounter{footnote}{0}

\begin{abstract}
  Loop quantum cosmology is a conflicted field in which exuberant
  claims of observability coexist with serious objections against the
  conceptual and physical viability of its current formulations. This
  contribution presents a non-technical case study of the recent claim that
  loop quantum cosmology might alleviate anomalies in observations of the
  cosmic microwave background.
\end{abstract}

\section{Introduction}

\begin{quote}
 ``Speculation is one thing, and as long as it remains speculation, one can
 understand it; but as soon as speculation takes on the actual form of ritual,
 one experiences a proper shock of just how foreign and strange this world
 is.'' \cite{Arendt}
\end{quote}

Quantum cosmology is a largely uncontrolled and speculative attempt to explain
the origin of structures that we see in the universe. It is uncontrolled
because we do not have a complete and consistent theory of quantum gravity
from which cosmological models could be obtained through meaningful
restrictions or approximations. It is speculative because we do not have
direct observational access to the Planck regime in which it is expected to be
relevant.

Nevertheless, quantum cosmology is important because extrapolations of known
physics and observations of the expanding universe indicate that matter once
had a density as large as the Planck density. Speculation is necessary because
it can suggest possible indirect effects that are implied by Planck-scale
physics but manifest themselves on more accessible scales. Speculation is
therefore able to guide potential new observations.

Speculation becomes a problem when it is based on assumptions that are
unmentioned, poorly justified or, in some cases, already ruled out. When this
happens, speculation is turned into a ritual followed by a group of
practitioners who continue to believe in their assumptions and ignore outside
criticism. The uncontrolled nature of quantum cosmology makes it particularly
susceptible to this danger.

A recent example is the claim \cite{Tango} that loop quantum cosmology may
alleviate various anomalies in observations of the cosmic microwave
background. Given the commonly accepted distance between quantum cosmology and
observations available at present, this claim, culminating in the statement
that ``these results illustrate that LQC has matured sufficiently to lead to
testable predictions,'' is surprising and deserves special scrutiny, all the
more so because a large number of conceptual and physical shortcomings have
been uncovered in loop quantum cosmology over the last few years. (All
citations from \cite{Tango} refer to its preprint version.)

On closer inspection, we encounter a strange world in the claims of
\cite{Tango} according to which, to mention just one obvious misjudgement,
``many of the specific technical points [of a recent critique of the methods
used in the analysis] were already addressed, e.g., in [42, 64, 65] and in the
Appendix of [66]'' even though references [42, 64, 65, 66] there were
published between nine and twelve years before the recent critique that they
are supposed to have addressed.

Clearly, the authors of \cite{Tango} have not sufficiently engaged with
relevant new criticism. There is therefore a danger that their work is based
on rituals.  Our analysis will, unfortunately, confirm this suspicion.  A
dedicated review of these claims is especially important because they have
been widely advertized, for instance in a press
release.\footnote{\url{https://news.psu.edu/story/626795/2020/07/29/research/cosmic-tango-between-very-small-and-very-large}}

The present paper provides a close reading of these claims and highlights
various shortcomings. It therefore serves as a case study of the complicated
interplay between quantum cosmological modeling and observations. A detailed
technical discussion of some of the underlying problems of current versions of
loop quantum cosmology has already been given in \cite{Claims}. The analysis
here is held at a non-technical level and is more broadly accessible. It
highlights conceptual problems that are relevant for \cite{Tango} but were not
considered in \cite{Claims}, for instance concerning suitable justifications
of initial states in a bouncing cosmological model.

Since the focus of our exposition is on quantum cosmology, other questions
that may well be relevant for an assessment of the claims of \cite{Tango} will
not be discussed.  These questions include a proper analysis of all
early-universe observables in addition to a select number of anomalies, and
the possible role of sub-Planckian unknowns such as further modifications of
gravity not considered in \cite{Tango} or uncertainties about the energy
contributions in early universe; see for instance
\cite{TensionBransDicke,TensionSnowmass,TensionRealm} for related examples and
reviews.

\section{Inconsistencies}

The model constructed in \cite{Tango} is based on several key assumptions made
elsewhere in the literature on loop quantum cosmology:
\begin{description}
\item[Assumption 1:] In loop quantum cosmology, there are quantum-geometry
  corrections that imply a bounce of the universe at about the Planck density.
\item[Assumption 2:] During quantum evolution through the bounce and long
  before and after, the quantum state of the universe remains sharply peaked
  as a function of the scale factor.
\item[Assumption 3:] Although the dynamics is modified and geometry is
  quantum, general covariance in its usual form is preserved on all scales,
  including the Planck scale.
\item[Assumption 4:] At the bounce, the spatial geometry and matter
  distribution is as homogeneous and isotropic as possible, restricted only by
  uncertainty relations for modes of perturbative inhomogeneity.
\end{description}
These assumptions are necessary for the model to work as it does. For
instance, the bounce (Assumption 1) and spatial symmetries (Assumption 4)
imply a certain cut-off in the primordial power spectrum that quickly turns
out to be useful in the context of anomalies in the cosmic microwave
background. A sharply peaked state (Assumption 2) and general covariance
(Assumption 3) make it possible to analyze the model by standard methods of
space-time physics, using line elements and well-known results from
cosmological perturbation theory \cite{CosmoPert}.

On closer inspection, however, it is hard to reconcile these assumptions with
established physics. Some of them are even mutually inconsistent. We will
first discuss these inconsistencies and then examine how the authors try to
justify their assumptions.

\subsection{Bounce}

In general relativity, there are well-known singularity theorems that prove
the existence of singularities in the future or the past, under certain
assumptions such as energy inequalities or the initial condition of a universe
expanding at one time \cite{HawkingEllis,Senovilla}. In their most general
form, these theorems do not require the specific dynamics of general
relativity but only use properties of Riemannian geometry, such as the
geodesic deviation equation. Loop quantum cosmology can avoid the big-bang
singularity and replace it by a bounce only if it violates at least one of the
assumptions of singularity theorems. Just having a theory with modified
dynamics compared with general relativity is not sufficient.

Loop quantum cosmology does not question that the universe is currently
expanding, thus obeying the same initial condition commonly used to infer the
singular big bang. (It is known that the specific modeling of an expanding
universe can be weakened so as to evade singularity theorems
\cite{NonSingSol}, but loop quantum cosmology does not avail itself of this
option.)  Moreover, as emphasized in \cite{Tango}, ``These qualitatively new
features arise without having to introduce matter that violates any of the
standard energy conditions.'' According to Assumption 3, space-time in loop
quantum cosmology remains generally covariant on all scales (described by a
suitable line element) and therefore obeys general properties of Riemannian
geometry. Loop quantum cosmology does modify the gravitational dynamics, but
this is not a required ingredient of singularity theorems. 

Since all the conditions of singularity theorems still hold, according to
explicit or implicit assumptions in \cite{Tango}, loop quantum cosmology
should be singular just like classical general relativity. How can it exhibit
a bounce at high density? For a detailed resolution of this conundrum, see
\cite{BlackHoleModels}.

\subsection{Peakedness}

Setting aside special dynamics such as the harmonic oscillator, quantum states
generically spread out and change in complicated ways as they evolve.  For a
macroscopic object, such as a heavy free particle, it takes longer for such
features to become significant than for a microscopic system, but they are
nevertheless present. 

In late-time cosmology, it is justified to assume that a simple homogeneous
and isotropic spatial geometry of a large region describes the dynamics very
well. At late times, a state in quantum cosmology may therefore be assumed to
remain sharply peaked because it describes a large region that is conceptually
analogous to a heavy free particle.

When such a state is extrapolated to the big bang, however, very long time
scales are involved. Is it still justified to assume that the state does not
change significantly and spread out during these times? Moreover, if the usual
assumption of a homogeneous and isotropic geometry is maintained for a
tractable description of quantum cosmology, such a region would have to be
chosen smaller and smaller as we approach the big bang, not only because of
the shrinking space of an expanding universe in time reverse but also, and
more importantly, because attractive gravity implies structure formation
within co-moving volumes. As the big bang is approached, a valid homogeneous
approximation of quantum cosmology more and more resembles quantum mechanics
of a microscopic object.

Because we do not know what a realistic geometry of our universe at the Planck
scale should be, we do not know how macroscopic it may still be considered to
be in a homogeneous approximation. The only indication is the
Belinskii--Khalatnikov--Lifshitz (BKL) scenario \cite{BKL} of generic
space-time properties near a spacelike singularity, which suggests that there
is no lower limit to the size of homogeneous regions in classical general
relativity. Quantum gravity is expected to modify the dynamics of general
relativity, but it is not known whether and at what scale it would be able to
prevent BKL-type behavior.

This result suggests that a microscopic description should be assumed in the
Planck regime. However, according to \cite{Tango}, ``One can now start with a
quantum state $\Psi(a,\phi)$ that is peaked on the classical dynamical
trajectory at a suitably late time when curvature is low, and evolve it back
in time towards the big bang using either the WDW equation or the LQC
evolution equation. Interestingly the wave function continues to remain
sharply peaked in both cases.''  Why should quantum cosmological dynamics be
such that it maintains a sharply peaked state even over long time scales that
include a phase in which the evolved object should be considered microscopic?

\subsection{Covariance}

According to Assumption 1, the bounce in loop quantum cosmology is supposed to
happen because quantum geometry modifies the classical dynamics: ``Of
particular interest is the area gap --- the first non-zero eigenvalue $\Delta$
of the area operator. It is a fundamental microscopic parameter of the theory
that then governs important macroscopic phenomena in LQC that lead, e.g., to
finite upper bounds for curvature.'' There is supposed to be a certain quantum
structure of space that leads, among other things, to a discrete area
spectrum. 

However, the description of perturbative inhomogeneity in \cite{Tango} or the
underlying \cite{AAN} assumes that this quantum geometry can be described by a
line element. To be sure, coefficients of the ``dressed'' metric in this line
element contain quantum corrections, but it has not been shown that a quantum
geometry with some discrete area spectrum can be described by any line element
at all, even at the Planck scale where discreteness is supposed to be
significant enough to change the dynamics of the classical theory. While
specific corrections in metric coefficients have been derived, the claim that
they should appear in a line element of some Riemannian geometry has not been
justified.

There are now several no-go results \cite{NonCovDressed,Disfig} that
demonstrate violations of covariance in regimes envisioned by the authors of
\cite{Tango}. A proof that Riemannian geometry and line elements can
nevertheless be used in their context would therefore require a detailed
discussion of how these no-go results can be circumvented. However, there is
no hint of such an attempt. It is worth noting that problems with covariance
have also occured in a different application of the same formalism to black
holes, in which several other physical problems were quickly found
\cite{DiracPoly,ExtendedPoly,TransCommAs,LoopISCO}. The space-time description
assumed in \cite{Tango} is therefore unreliable.

Given the uncertain status of what structure space-time geometry should have
in the presence of modifications from loop quantum cosmology, the meaning of
``finite upper bounds for curvature'' that are supposed to be implied by the
discrete area spectrum is unclear.

\subsection{Symmetry}

Spatial homogeneity and isotropy are assumed at various places in
\cite{Tango}. First, in older papers referred to for justifications of some
claims, quantum evolution is numerically computed for states of an exactly
homogeneous and isotropic geometry, using methods from quantum
cosmology. Secondly, states for inhomogeneous matter perturbations on such a
background are assumed to preserve the symmetry as much as possible while
obeying uncertainty relations.

\subsubsection{Background}

An attempt is made in \cite{Tango} to justify these assumptions: ``On the
issue of simplicity of the LQC description, we note that in the 1980s it was
often assumed that the early universe is irregular at all scales and therefore
quite far from being as simple as is currently assumed at the onset of
inflation. Yet now observations support the premise that the early universe is
exceedingly simple in that it is well modeled by a FLRW spacetime with first
order cosmological perturbations. Therefore, although a priori one can
envisage very complicated quantum geometries, it is far from being clear that
they are in fact realized in the Planck regime.''

However, this attempted justification is invalid because the scale probed by
early-universe observations referred to in this statement is vastly different
from the Planck scale in which the outcome is applied. Cosmic inflation was
proposed precisely to address, among other things, the homogeneity problem
even if the initial state may be much less regular, and inflation is still
used in most models of loop quantum cosmology. In such scenarios, large-scale
homogeneity at later times does not show in any way that the universe must
have been homogeneous at the Planck scale.

The assumed homogeneity is also in conflict with the claimed discrete
structure of space that, according to Assumption 1, might imply a bounce. The
authors of \cite{Tango} never address the relevant question of how their
symmetry assumptions can be reconciled with a discrete geometry on ultraviolet
scales that is supposed to imply all the claimed effects. (In other words, the
authors ignore the trans-Planckian problem of inflationary cosmology
\cite{TransPlanck,TransPlanck2,TransPlanck3}.) Of course, suitable
superpositions of discrete states may lead to a continuum of expectation
values even of operators that have a discrete spectrum. But if this argument
were used as a possible explanation of homogeneity, it would put in doubt the
strong emphasis on a single eigenvalue $\Delta$ of the area spectrum that is
claimed to govern the new dynamics.

\subsubsection{Perturbations}

The authors of \cite{Tango} use symmetry assumptions not only for the
background on which inhomogeneous modes evolve but also for the modes
themselves, described perturbatively. These modes cannot be exactly symmetric
because they are subject to uncertainty relations and therefore have, at
least, non-zero fluctuations even if their expectation values may be zero. It
is claimed that these modes should be as symmetric as possible in the bounce
phase, that is, have zero expectation values and fluctuations such that they
saturate uncertainty relations.

The homogeneity assumption for perturbations is motivated by Penrose's Weyl
curvature hypothesis \cite{PenroseEntropy}, introduced in \cite{Tango} by
``Finally, the principle that determines the quantum state $\Psi(Q,\phi)$ of
scalar modes involves a quantum generalization of Penrose's Weyl curvature
hypothesis in the Planck regime near the bounce, which physically corresponds
to requiring that the state should be `as isotropic and homogeneous in the
Planck regime, as the Heisenberg uncertainty principle allows'.''  However,
conceptually there is a significant difference between these two
proposals. Penrose's hypothesis was made in a big-bang setting in which the
initial state (close to the big-bang singularity) was to be restricted by
geometrical considerations. As always, one may question the specific
motivation for a certain choice of initial states, but there are no physical
objections provided general conditions (such as uncertainty relations) are
obeyed.

The symmetry assumption employed for matter perturbations in \cite{Tango} is
of a very different nature. It is used to determine an initial state only of
our current expanding phase of the universe, but it is set in a bounce model
with a pre-history before the big bang. The symmetry assumption for matter
perturbations is therefore a final condition for the collapse phase and
violates determinism. Potential violations of deterministic behavior have
indeed been derived in models of loop quantum cosmology in the form of
signature change at high density \cite{Action,SigChange,SigImpl}. However,
\cite{Tango} does not make use of this option, which would in fact be in
conflict with the line element they assume to formulate a wave equation for
cosmological perturbations. Moreover, the derived versions of signature change
would set the beginning of the Lorentzian expanding branch of the universe
later than assumed in \cite{Tango}, at the very end of the bounce phase in
which modifications from loop quantum cosmology subside.

Setting aside the question of determinism, a restriction of the state during
the bounce phase, a transitory stage, is not an initial condition but rather
an assumption about the state that preceding collapse may have led to.  It is
then questionable that gravitational collapse of a preceding inhomogeneous
universe should lead to a bounce state that is as homogeneous as possible,
respecting uncertainty relations. In this scenario, collapse of a preceding
universe is supposed to have led to a very homogeneous state at the Planck
density of more than one trillion solar masses in a proton-sized region. Given
the inherently unstable nature of gravitational collapse, one would rather
expect that any slightly overdense region in a collapsing universe would
quickly become denser and magnify initial inhomogeneities that had been
present when collapse commenced.

It is hard to see how collapse could, instead, lead to a state that is as
homogeneous as possible. Such an assumption would at least require a dedicated
justification, in particular because it directly implies the crucial features
of a new scale in loop quantum cosmology that is then used to alleviate
anomalies. Unfortunately, no attempted justification can be found in
\cite{Tango}.

On closer inspection, the arguments given in \cite{Tango} are even
circular. The very setup that leads the authors to their formulation of a
homogeneous initial condition already assumes near homogeneity of a collapsing
universe: The specific statement in \cite{Tango} refers to ``the Planck regime
near the bounce'' and implicitly assumes that (in the Riemannian geometry
according to Assumption 3) there is a time coordinate such that ``the bounce''
happens everywhere within a large region at the same time. However, if the
collapsing geometry is inhomogeneous, overdense regions will become denser
during collapse and reach the Planck density earlier than their
neighbors. Once they bounce and start expanding, it is not obvious that a
simple near-homogeneous slicing with a uniform bounce time still exists.
(This process suggests a multiverse rather than a single nearly homogeneous
universe \cite{Multiverse}.)

There is no unique bounce time in this picture, and therefore an implicit
assumption (a meaningful ``near the bounce'') used in the condition of initial
homogeneity is unphysical. Crucial statements such as ``In our LQC model, the
physical principles used to select the background quantum geometry imply that
the corresponding $\Lambda$CDM universe has undergone approximately 141
e-folds of expansion since the quantum bounce until today.'' therefore remain
unjustified.

\subsection{Attempted justifications}

The authors of \cite{Tango} do realize that some of these assumptions should
be justified, while they are apparently unaware of additional implicit
assumptions that they do not mention. 
\begin{enumerate}
\item The bounce is justified by quantum-geometry corrections from loop quantum
gravity, but the conflict with singularity theorems has apparently gone
unnoticed.
\item The peakedness of states is justified by referring to detailed numerical
  studies of evolving states in loop quantum cosmology. However, the authors
  fail to notice that these studies implicitly assume that the universe is
  still macroscopic (large-scale homogeneity) even in the Planck regime,
  ignoring structure formation within co-moving regions in a collapsing
  universe as well as BKL-type behavior.
\item Covariance is justified only vaguely by referring to wave equations for
perturbations on a background, without asking the relevant question of whether
modified perturbation and background equations can still be consistent with a
single metric that is being perturbed.
\item In their discussion of initial conditions for perturbations, the authors
  seem to be unaware of problems posed by the pre-history of a collapsing
  universe.
\end{enumerate}

\section{A brief engagement with previous critique}
\label{s:Engage}

It is instructive to analyze the brief but rapid-fire response given in
\cite{Tango} to previous criticism of certain claims in loop quantum cosmology
\cite{Claims}. As already mentioned in the Introduction, the authors state
that ``Many of the specific technical points were already addressed, e.g., in
[42, 64, 65] and in the Appendix of [66].''  However, the references provided
try (but fail \cite{RecallComment,Casimir}) to address an older issue, cosmic
forgetfulness \cite{BeforeBB,Harmonic}, that does not play a major role in the
recent discussion of \cite{Claims}. (Since the publication of these older
papers, cosmic forgetfulness has been strengthened to signature change.)

\subsection{Effective descriptions}

The authors of \cite{Tango} go on and state that ``First, although `effective
equations' are often used in LQC, conceptually they are on a very different
footing from those used in effective field theories: One does not integrate
out the UV modes of cosmological perturbations. The term `effective' is used
in a different sense in LQC: these equations carry some of the leading-order
information contained in sharply peaked quantum FLRW geometries
$\Psi(a,\phi)$.'' (Note that the authors are hedging their statement by
correctly saying that ``these equations carry {\em some} of the leading-order
information,'' emphasis added. The fact that they carry only some but not all
of the leading-order information is a problem in itself that will not be
discussed here; see \cite{ROPP,Claims} for details.)

The admission that ``the term `effective' is used in a different sense in LQC''
does not make this formalism more meaningful. It is in fact one of the major
problems in current realizations of the framework of loop quantum
cosmology. Equations of loop quantum cosmology are used on vastly different
scales, in the Planck regime to analyze the possibility of a bounce, and at
low curvature to justify a peaked late-time state. A suitable effective theory
(not in the sense used, according to \cite{Tango}, in loop quantum cosmology)
would be needed to determine how parameters of the model may change by
infrared or ultraviolet renormalization. The authors are simply assuming that
a single effective theory without any running parameters can be used to
describe the quantum universe on a vast range of scales. There is no
justification for this assumption.

In addition, the authors refer not only to different scales in cosmology but
also to different geometries, including those of black-hole horizons: When
they justify certain parameter choices for the dynamics of loop quantum
cosmology, they make statements such as ``The eigenvalues of [the area
operator] $\hat{A}_S$ are discrete in all $\gamma$-sectors.  But their
numerical values are proportional to [the Barbero--Immirzi parameter (a
quantization ambiguity)] $\gamma$ and vary from one $\gamma$ sector to
another.'', ``In LQG, a direct measurement of eigenvalues of geometric
operators would determine $\gamma$. But of course such a measurement is far
beyond the current technological limits.'' and ``Specifically, in LQG the
number of microstates of a black hole horizon grows exponentially with the
area, whence one knows that the entropy is proportional to the horizon
area. But the proportionality factor depends on the value of
$\gamma$. Therefore if one requires that the leading term in the statistical
mechanical entropy of a spherical black hole should be given by the
Bekenstein-Hawking formula $S = A/4\ell_{\rm Pl}^2$, one determines $\gamma$
and thus the LQG sector Nature prefers.'' 

These statements refer to at least three different regimes of some underlying
theory of loop quantum gravity: direct microscopic measurements of
eigenvalues, macroscopic black-hole horizons, and the entire universe at
various densities. It is simply assumed that the same value of $\gamma$ (as
well as other parameters) may be used in all these situations without suitable
renormalization. The only justification attempted for this assumption is the
statement that ``the term `effective' is used in a different sense in LQC''
which ignores the physical reasons for standard ingredients in effective
theories.

The possibility of running is also ignored in statements such as ``In IV C, we
will show that the interplay between LQC and observations is a 2-way bridge,
in that the CMB observations can also be used to constrain the value of the
area gap $\Delta$, the most important of fundamental microscopic parameters of
LQG.'', ``As we saw in section III, the area gap $\Delta$ is the key
microscopic parameter that determines values of important new, macroscopic
observables such as the matter density and the curvature at the bounce. Its
specific value, $\Delta = 5.17 \ell_{\rm Pl}^2$, is determined by the
statistical mechanical calculation of the black hole entropy in loop quantum
gravity (see, e.g., [55, 56, 59, 60]).'' and ``Clearly, the value
$\Delta\approx 5.17\ell_{\rm Pl}^2$ chosen in Sec. III C and used in this
paper, is within 68\% (1$\sigma$) confidence level of the constraint obtained
from Planck 2018. This not only indicates a synergy between the fundamental
theoretical considerations and observational data, but also provides internal
consistency of the LQC model.''

\subsection{Covariance}

In their reply to \cite{Claims}, the authors state that ``As we will see in
Section III B, equations satisfied by the cosmological perturbations are
indeed covariant.'' In Section III.B, however, less than half a sentence is
devoted to this important question: ``Note also that the equation is covariant
w.r.t.\ [the dressed metric] $\tilde{g}_{ab}$ and $\tilde{g}_{ab}$ rapidly
tends to the classical FLRW metric of GR outside the Planck regime.'' This
attempted justification of covariance apparently refers to the equation
$(\tilde{\Box}+\tilde{U}/\tilde{a}^2)\hat{Q}=0$ for modes $\hat{Q}$, with
certain functions $\tilde{U}$ and $\tilde{a}$ of the background scale factor.
But nowhere do the authors address the crucial question of whether background
$\tilde{g}_{ab}$, which defines the d'Alembertian $\tilde{\Box}$, and
perturbation $\hat{Q}$ can after modifications still be obtained from a single
covariant metric (obeying the tensor-transformation law) as in the underlying
classical theory.

The authors' statement about covariance refers only to transformations of the
perturbative mode, $\hat{Q}$, and therefore to small inhomogeneous coordinate
changes that preserve the perturbative nature. In addition, potentially large
homogeneous transformations of the background time coordinate, such as
transforming from proper time to conformal time, are relevant in cosmological
models of perturbative inhomogeneity.  The usual curvature perturbations are
not invariant with respect to these transformations \cite{Stewart}.  The
authors' statements have nothing to say about the question of whether their
model of modified perturbation equations is covariant with respect to large
transformations of background time. The failure of covariance in the
underlying construction has been shown in \cite{NonCovDressed}.

\subsection{Symmetry}

We have already addressed the authors' erroneous view that observations can be
used to justify near homogeneity in the Planck regime. The authors finally
state that ``Nonetheless, one should keep in mind that, as in other approaches
to quantum cosmology, in LQC the starting point is the symmetry reduced,
cosmological sector of GR. Difference from the Wheeler-DeWitt theory is that
one follows the same systematic procedure in this sector as one does in full
LQG. But the much more difficult and fundamental issue of systematically
deriving LQC from full LQG is still open mainly because dynamics of full LQG
itself is still a subject of active investigation.'' Here, they are attempting
to construct a false binary choice between exactly isotropic models of
symmetry-reduced geometries on one hand, and calculations in full loop quantum
gravity without any symmetry assumptions on the other. If this choice were
correct, one might as well give up because exactly isotropic models are
unrealistic, and the full theory is untractable.

What the authors are missing is a proper effective theory that does not only
amend isotropic equations by certain leading-order corrections, but also tries
to go beyond strictly isotropic models by parameterizing all the ignorance in
parameter choices implied by the intractable nature of full loop quantum
gravity. Without such an effective description, which does not require direct
calculations in the full theory and is therefore feasible but would not be as
simple as the authors assume, no observational claims can be reliable.

\section{Conclusions}

How can we reconcile the claim that ``these results illustrate that LQC has
matured sufficiently to lead to testable predictions'' \cite{Tango} with the
availability of several serious and independent concerns that have shown in
recent years how loop quantum cosmology, as it is commonly practiced, has
overlooked a large number of important conceptual and physical requirements?
In the present paper, we have provided detailed evidence to show that
\cite{Tango} has merely ignored or insufficiently addressed relevant
criticism. Moreover, the main new claims of \cite{Tango} are implied rather
directly by specific assumptions that remain unjustified. We summarize these
observations in this concluding section.

In their quest to show that loop quantum cosmology naturally resolves
anomalies in observations of the cosmic microwave background and therefore
makes testable predictions, the authors of \cite{Tango} have used or
introduced several conceptually distinct assumptions. Some of these
assumptions, including those about bounces and covariance, are questionable
within loop quantum cosmology. Others, for instance about free parameters,
rely on an oversimplified presentation of loop quantum cosmology as some
special version of an effective theory that could be used to describe the
dynamics of quantum gravity on a vast range of scales, including the Planck
regime, without any running parameters.  

Yet another set of assumptions, referring to the state of perturbations and
their spatial homogeneity, is independent of loop quantum cosmology but has
been packaged with the other assumptions in a way that gives the erroneous
impression of a single coherent theory. Moreover, these assumptions are
physically questionable because they implicitly make strong and unrealistic
claims about the generic final state of a collapsing universe. We are
observing a single universe which might perhaps have emerged from a special
version of preceding collapse as envisioned by the authors. But by simply
formulating the desired behavior as an assumption without addressing a
possible relationship with a pre-history, the authors hide its restrictive
nature and ultimately fail to explain the initial state of cosmological
perturbations and the observed microwave background.

The authors of \cite{Tango} are aware of previous criticism and make a
half-hearted attempt to address it. In Section~\ref{s:Engage} we have seen how
inadequate their response is. To summarize, the cited papers that supposedly
addressed ``many of the specific technical points'' made in \cite{Claims} had
been published between nine and twelve years before this recent critique. The
crucial question of general covariance in models of quantum gravity, discussed
from different viewpoints for instance in
\cite{LorentzFineTuning,SmallLorentzViol}, is acknowledged in \cite{Tango}
only by the misleading ``as we will see in Section III B, equations satisfied
by the cosmological perturbations are indeed covariant'' to annnounce a brief
statement ``note also that the equation is covariant w.r.t.\ [the dressed
metric] $\tilde{g}_{ab}$'' that reflects a misunderstanding of the issue of
covariance in the setting of modified perturbation equations. (Not only
perturbations but also the background must be included in a covariance
analysis \cite{NonCovDressed}.)

Notably, there are also points of critique that the authors of \cite{Tango} do
not bother to address. An important example, in addition to the prevalence of
quantization ambiguities, is the implicit assumption in current formulations
of loop quantum cosmology that the universe remains large-scale homogeneous
(over at least several hundred Planck lengths) even at the Planck density. As
explained in \cite{Claims}, this assumption is related to the incorrect
implementation of effective descriptions in loop quantum cosmology. Without
this unjustified and unmentioned assumption, the authors of \cite{Tango} would
not even be able to formulate their restrictive condition that cosmological
perturbations be as homogeneous as possible at the bounce. This assumption
ultimately leads to a suppression of power on large scales of the cosmic
microwave background and allows claims about resolved anomalies. However, this
assumption is not only unjustified, it is also based on another and implicit
assumption that had already been ruled out.

Therefore, the results of \cite{Tango} rely on several major assumptions, made
explicitly or implicitly, that turn out to be unjustified. The authors' claims
therefore go well beyond the usual level of speculation that is common (and
unavoidable) in quantum cosmology. Their response to objections that have been
published during the last few years is inadequate in some cases and
non-existent in others.  Ashtekar et al.\ have created a cosmic tangle of
rituals that they no longer wish to be questioned.

\section*{Acknowledgements}

This work was supported in part by NSF grant PHY-1912168.

%\bibliographystyle{../preprint}
%\bibliography{../Bib/QuantGra.bib}

\end{document}